\documentclass[12pt]{article}
\usepackage{amsmath,epsfig,amssymb,eufrak}

\date{}

\newcommand{\dds}{\stackrel{\leftrightarrow}{D}}

\input macros.sty   
\input eqalign.sty  

\begin{document}

\begin{titlepage}

\title{
  {\vspace{-0cm} \normalsize
  \hfill \parbox{40mm}{DESY/03-028\\
                       SFB/CPP-4\\March 2003}}\\[30mm]
Continuous external momenta in\\non-perturbative
lattice simulations: \\a computation of renormalization factors } 
  \author{M.\ Guagnelli$^{a}$, K.\ Jansen$^{b}$, F.\ Palombi$^{a}$,\\ 
   R.\ Petronzio$^{a}$, A.\ Shindler$^{b}$ and I.\ Wetzorke$^{b}$\\
\\
   {Zeuthen-Rome (ZeRo) Collaboration}\\
\\ 
  {\small $^a$ Dipartimento di Fisica, Universit\`a di Roma 
        {\em Tor Vergata}}\\ 
  {\small and INFN, Sezione di Roma II,} \\
  {\small Via della Ricerca Scientifica 1, I-00133 Rome, Italy} \\
  {\small $^b$ NIC/DESY Zeuthen, Platanenallee 6, D-15738 Zeuthen,
   Germany}\\
}

\maketitle

\begin{abstract}
We discuss the usage of continuous external momenta 
for computing renormalization 
factors as needed to  renormalize operator matrix elements. 
These kind of external momenta are encoded in special boundary conditions for
the fermion fields. The method allows to compute certain renormalization
factors on the lattice that would have been very difficult, if not impossible,
to compute with standard methods. As a result we give the 
renormalization group invariant
step scaling function for a twist-2  operator corresponding to the average
momentum of non-singlet quark densities.
\vspace{0.8 cm}
\noindent
\end{abstract}

\end{titlepage}

\section{Introduction}

One of the most important contributions lattice gauge theory can provide to test
QCD and to interpret experimental data, is the computation of
non-perturbatively renormalized matrix elements. The example we are 
interested in here are matrix elements that are connected to moments
of parton distribution functions as can be extracted from global 
fits to experimental data in deep inelastic scattering. 
The matrix elements are related themselves to certain 
operator expectation values which are accessible from lattice
simulations. 

When considering operators that contain derivatives, 
a saturation with external momenta is needed to perform the 
necessary contractions. This 
applies for computing the renormalization constants as
well as matrix elements of such operators. 
On an euclidean lattice, the standard momenta are quantized in 
units of $2\pi a/L$ with $a$ the lattice spacing and $L$ the linear
extent of a lattice of physical size $L^4$. 
Unfortunately, the introduction of a 
momentum in a numerical simulation 
leads often to either large lattice artefacts or to large   
statistical uncertainties or to both such that reliable 
measurements of such quantities become difficult.

In this paper we will demonstrate how the introduction of special 
boundary conditions of the fermion fields, first advocated
in \cite{campi_bordo}, can help in such a 
situation. The boundary conditions lead to continuous values
of external momenta $a\theta/L$ where $\theta$ varies 
continuously in the range   
$0 \le \theta \le 2\pi$ giving a large flexibility of a 
momentum definition in lattice simulations.
In particular, the minimal value of a lattice momentum 
can be chosen smaller than in the standard set-up which might lead 
to smaller lattice artefacts. 
As we will see in the example discussed below, the use of these 
kind of continuous external 
momenta will allow us to compute renormalization constants reliably on 
the lattice 
for which measurements were very difficult in the standard set-up.
It might be,
however, that the method is much more general and goes beyond
the application investigated here.
 
The physical example we will discuss in this paper  
is the calculation 
of renormalization constants as needed for twist-2, non-singlet
quark operators in the framework of computing moments of
parton distribution functions in deep inelastic scattering. 
The interest in this calculation is twofold. First, we want 
to demonstrate the feasibility of using continuous
momenta in a lattice computation. 

Second, the success of this
demonstration will allow us to correct a small mismatch of our
earlier work:
In a series of papers 
\cite{ref:perturbative,ref:non-pert,ref:universal,ref:invariant,ref:matrixele}
we have demonstrated that the Schr\"odinger functional (SF) formalism
\cite{sf_fond} can be used to compute non-perturbatively moments 
of parton distribution functions (for summaries of the results, see
\cite{karl_osaka,andrea_elba,karl_genua}). 
A problem that arises naturally in lattice computations
of such quantities is that
for a given continuum operator there is not a unique representation
of that operator on the lattice.
In fact, in the old work cited above, the operator used to
compute the
matrix element and the operator used to compute 
the corresponding renormalization constant were in two different
representations.
Although in ref.~\cite{ref:matrixele}
we argued that this mismatch only leads to a small,
negligible error, the situation
remains rather unsatisfactory, as we want a fully
non-perturbative way to evaluate the physical matrix element without
systematic uncertainties.
The usage of the continuous external momenta will allow us  
to eliminate this small systematic uncertainty. 
In particular, we will show that we can  
compute  
the scale dependent renormalization constant and the corresponding
(ultra-violet) invariant step scaling function for a particular 
lattice representation which is 
needed to perform the correct renormalization for the physical 
matrix element. In this way, we will be able to provide a 
non-perturbative lattice computation of the average momentum 
carried by the quarks in a hadron.

\section{Basic definitions and choice of boundary conditions}

A bare, local operator $\mathcal{O}^\mathrm{bare}$ as considered here,
is renormalized multiplicatively by a scale $\mu$ dependent
renormalization constant $Z_{\cO}(\mu)$.                     
The renormalized operator $\cO^R(\mu)$ is then given by 
\be
\cO^R(\mu) = Z_{\cO}^{-1}(\mu) \cO^{\rm bare}\; .
\ee
The evolution of $\cO$ from a renormalization scale $\mu_1$ to $\mu_2$ 
is described by the continuum step scaling function
$\sigma_{Z_{\cO}}$
that can formally be written as
\be
\sigma_{Z_{\cO}}(\mu_1/\mu_2) = \frac{Z_{\cO}(\mu_2)}{Z_{\cO}(\mu_1)}\; .
\label{firstssf}
\ee
The individual $Z$'s in eq.~(\ref{firstssf}) are only well-defined 
within a given regularization scheme and would be divergent when the 
regularization is removed. 
The step scaling function, however, is a well-defined quantity 
even in this limit. 
In the following we will give a rigorous definition of the renormalization
constants and of the step scaling function using the lattice regularization
in the Schr\"odinger functional (SF) scheme.

The SF scheme is based on the formulation
of QCD in a finite space-time volume of size $L^3 \cdot T$.
In this paper we will always use $L=T$. 
In this scheme a change of renormalization scale
amounts to a change of the box size $L$ at fixed bare parameters.
By considering a sequence of pairs of volumes with sizes $L$ and $sL$, one
can study the evolution of a given local operator under repeated changes
of the scale by a factor of $s$. Effectively in this way 
one builds up a non-perturbative renormalization group.
To be specific, we will from now on work on a euclidean lattice as regulator. 

The fermion and gauge fields on the lattice are defined in the standard way, 
fulfilling SF boundary 
conditions, as detailed in 
ref.~\cite{sf_fond}. Besides these special boundary conditions 
in time direction, we will impose generalized 
boundary conditions in the spatial directions (denoted by $\hat{k}$)
for the fermion fields $\psi(x)$ \cite{campi_bordo}
\begin{equation}
\psi(x+L\hat{k})=e^{i\theta_k}\psi(x),\;\; 
\bar{\psi}(x+L\hat{k})=e^{-i\theta_k}\bar{\psi}(x),\; k=1,\; 2,\; 3,
\label{theta_bc}
\end{equation}
while the gauge fields $U(x,k)$ are
chosen to be periodic in the space directions.
The values of $\theta_k$ can be chosen in the 
interval $0 \le \theta_k \le 2\pi$.
For $\theta_k=0$ we obtain periodic and for $\theta_k=\pi$ we get
anti-periodic boundary conditions. 

The generalized boundary conditions of eq.~(\ref{theta_bc}) 
can  
be implemented in the definition of the gauge covariant lattice 
derivatives: 
\begin{eqnarray}
\nabla_\mu\psi(x) & = & 
      \frac{1}{a}\left[ \lambda_\mu U(x,\mu)\psi(x+a\hat{\mu})-\psi(x)\right]
                 \nonumber \\
\nabla_\mu^{*}\psi(x) & = & 
  \frac{1}{a}\left[\psi(x) - \lambda_\mu^{*}U(x-a\hat{\mu},\mu)\psi(x-a\hat{\mu})
             \right]
\label{derivatives}
\end{eqnarray}
with $\mu=(k,4)$, 
\begin{equation}
\lambda_\mu = e^{ia\theta_\mu/L},\;\;  
              0 \le \theta_k \le 2\pi ,\;\;
              \theta_4=0,
\label{lambdas}
\end{equation}
and $a$ the lattice spacing. From a technical point of view the fields 
$\psi(x)$ and $\bar\psi(x)$ are then
implemented with the usual periodic boundary conditions in the space directions
utilizing this generalized definition of the covariant derivative of 
eq.~(\ref{derivatives}).

The crucial observation \cite{campi_bordo} is that 
the factor $e^{ia\theta_k/L}$ can be interpreted as an {\em external}
momentum with the intriguing property that it can assume   
continuous values, in contrast to the 
standard, quantized lattice momenta that assumes values 
in units of $2\pi a/L$ only.

In order to explore the flexibility of a momentum definition given 
by the generalized boundary conditions 
in eq.~(\ref{theta_bc}) we have concentrated in this
work on the twist-2, non-singlet (quark) operator. 
This amounts to consider operators of the form (the flavor
structure is specified by the Pauli matrix $\tau^3$)
\be
{\mathcal O}_{\mu \nu}(x) = \frac{1}{4}\bar\psi (x) \gamma_{\{\mu}
\lrD_{\nu\}}\frac{1}{2} \tau^3\psi (x) - \delta_{\mu \nu} \cdot {\rm trace~terms}\; .
\label{eq:ope1_tr}
\ee
where $\{\cdots\}$ means symmetrization on the Lorentz indices and
\be
\lrD_{\mu} = \rD_{\mu} - \lD_{\mu}; \qquad  D_{\mu} = 
\frac{1}{2}[\nabla_\mu + \nabla_\mu^{*}]\; .
\ee
There are
two representations of such a non-singlet operator on the lattice
\cite{h4_first}. The first representation takes $\mu\ne\nu$ 
whereas the second uses $\mu=\nu$. The precise definitions of the
operators used here are 
\begin{equation}
{\cal O}_{12}(x) = \frac{1}{4}
       \bar\psi(x) \gamma_{\{1} \dds_{2\}}\frac{1}{2} \tau^3 \psi(x)
\label{O12}
\end{equation}
and
\be
\cO_{44}(x) = \frac{1}{2} \bar\psi(x) \Big[ \gamma_4 \lrD_4 - \frac{1}{3}
\sum_{k=1}^3 \gamma_k \lrD_k \Big] \frac{\tau^3}{2} \psi(x)\; .
\label{O44}
\ee
In both cases an external momentum has to be supplied 
to compute their renormalization constants. We will realize 
these momenta 
with non-vanishing values of $\theta_k$, implicitly given
through the covariant derivatives in the fermion action and the 
operators in eqs.~(\ref{O12}), (\ref{O44}). 
The precise choice we adopt here is 
\begin{equation}
\theta \equiv \theta_1\ne 0, \; \theta_2=\theta_3=0\; .
\label{thetachoice}
\end{equation}

In order to make the discussion in the following self-consistent, 
let us recall the definition of bare correlation functions in the 
SF scheme \cite{ref:non-pert} as needed to compute appropriate 
renormalization factors. The correlation function of a given operator
$\cO$ is given by
\begin{equation}
  f_\cO(x_4/L,\theta)=-{a^6\over{L^3}}\sum_{\bf x,y,z}
  \langle \cO_{\mu\nu}(x)\
  \zetabar({\bf y})\Gamma\frac{1}{2}\tau^3\zeta({\bf z})\rangle\; ,
  \label{fO}
\end{equation}
where $\Gamma$ is a Dirac matrix that in our case is $\gamma_2$ for
the $\cO_{12}$ operator, while it is $\gamma_1$ for $\cO_{44}$.
In eq.~(\ref{fO}) $\zeta$ and $\zetabar$ are classical boundary 
fields at $x_4=0$ and we sum over space.                         
In order to normalize such correlation functions properly, we also need
the ``boundary to boundary'' correlation function $f_1$,
\begin{equation}
  f_1(\theta) =-{{a^{12}}\over{L^6}}
  \sum_{\bf u,v,y,z}
  \langle\zetabar'({\bf u})
  \gamma_5\frac{1}{2}\tau^a\zeta'({\bf v})
  \zetabar({\bf y})\gamma_5\frac{1}{2}\tau^a\zeta({\bf z})\rangle\; .
  \label{f1}
\end{equation}
Note that the classical boundary source fields 
$\zeta,\zetabar$  at $x_4=0$ and  $\zeta',\zetabar'$ at $x_4=L$
are renormalized multiplicatively with a common renormalization
constant $\zzeta$~\cite{StefanII,pert_1}.
The momentum ($\theta$) dependence of the 
correlation functions $f_{\cO}$ in eq.~(\ref{fO}) and $f_1$ 
in eq.~(\ref{f1}) is again implicitly given through the 
definition of the covariant derivative eq.~(\ref{derivatives})
that appears in the fermion action and in the operator considered. 

It is instructive to look at the $\theta$-dependence of the correlation
functions $f_{\cO_{44}}$ and $f_{\cO_{12}}$
for fixed values of $x_4/L=1/2$ and $x_4/L=1/4$ already at tree-level.   
\begin{figure}
\vspace{-1cm}
\begin{center}
\psfig{file=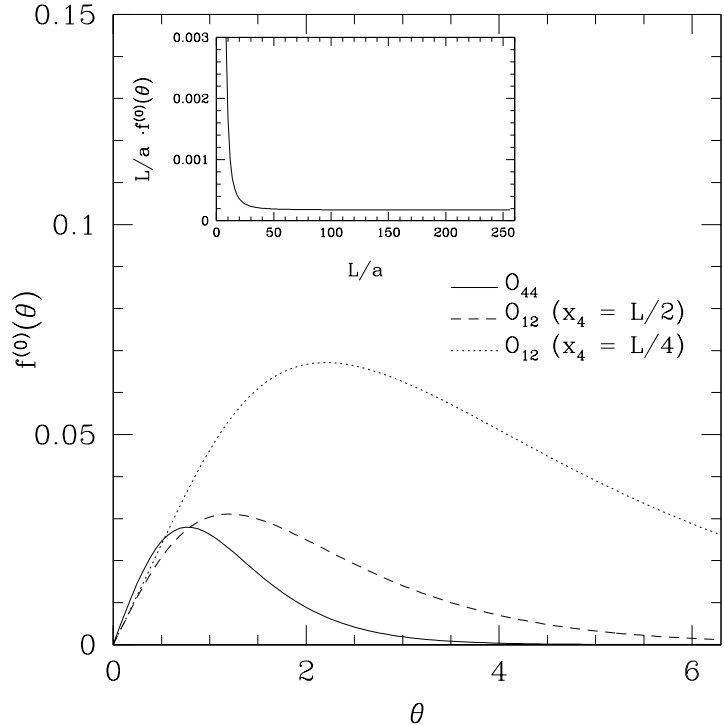, width=10cm}
\end{center}
\caption{ \label{fig:tree}
$\theta$-dependence of $f_{\cO}$ at tree-level. For the $\cO_{44}$ 
representation
the tree-level values at $x_4 = L/2$ and $x_4 = L/4$ are the same.
}
\end{figure}
Fig.~\ref{fig:tree} shows the tree-level correlation functions
$f_{\cO}^{(0)}$. It can be observed that the signal strongly 
depends on the chosen value of $\theta$, the definition of the
operator and the choice of $x_4/L$. 
For small values of $\theta$ all correlation functions show a linear
behavior in $\theta$ and vanish at $\theta=0$. 
In contrast to the correlation function of the operator $\cO_{12}$, 
the correlator for $\cO_{44}$, which does not depend on $x_4/L$ at tree-level,
becomes very small for values 
of, say, $\theta > 4.5$, see the inlay in fig.~\ref{fig:tree}. 
This implies that for a real simulation there is a danger that for these
$\theta$-values the signal may become very small, too. In addition,  
it seems that for these values of $\theta$, due to large lattice artifacts
(cf. sect. 2.2) it may become hard to extract even the one-loop 
anomalous dimension. 
It should be concluded therefore that the choice of $\theta$ in a simulation 
needs an optimization procedure. 
Before we can start a discussion of how to perform this optimization  
we first want to give here 
the 
renormalization conditions we have used in order to determine 
the renormalization factors which are defined as follows.
\begin{equation} \meqalign{
  \zco(a/L,x_4/L,\theta,\theta')~&=&~c~f_\cO(a/L,x_4/L,\theta)/
                            \sqrt{f_1(a/L,\theta')} \cr
  \zbarco(a/L,x_4/L,\theta)~&=&~\bar c~f_\cO(a/L,x_4/L,\theta)\;  
}
  \label{ZO}
\end{equation}
where the correlation functions are being evaluated at 
vanishing quark mass $\mq=0$ and for vanishing
boundary gauge and fermion fields. 
The renormalization conditions
are chosen such  
that $\zco=1~ {\rm{and}}~ \zbarco = 1$ at tree-level of 
perturbation theory at a fixed scale 
$\mu=1/L$ \cite{ref:perturbative,ref:non-pert}
while the external parameters $\theta$, $\theta'$ and $x_4/L$
can be varied. 
From these conditions we obtain
\begin{equation}\meqalign{
  c=\frac{\sqrt{f_1^{(0)}}}{f_\cO^{(0)}}\; , \;\;
  \bar c=\frac{1}{f_\cO^{(0)}}\; .
}
\end{equation}
The quantities $c~ {\rm{and}}~ \bar c$ are the expressions of the 
correlation functions in eq.~(\ref{fO}) and in eq.~(\ref{f1}) 
at tree-level with the corresponding
arguments. Note that the actual renormalization constant needed is 
$\zo$ whereas $\zbaro$ is only an auxiliary quantity that, however,  
proved useful in our previous work \cite{ref:non-pert}.
Note that in the definition of the renormalization constants in 
eq.~(\ref{ZO}) we leave the freedom to choose $\theta\ne\theta'$ 
for $f_\cO$ and $f_1$. 
It is important to stress here that a different choice of $x_4/L$, $\theta$ and
$\theta'$ defines a different renormalization scheme.
In this paper we have considered two families of renormalization schemes,
\begin{itemize}
\item {\rm scheme A} $\qquad \qquad \theta' = \theta$
\item {\rm scheme B} $\qquad \qquad \theta' = 0$\; .
\end{itemize}
Both schemes are studied in the following with respect to 
analyzing the convergence of perturbation theory,
the size of cutoff effects and the signal to noise ratio.
In these studies 
the external parameters $\theta$ and $x_4/L$ are tuned to 
optimize these criteria. 

\subsection{Convergence of perturbation theory}

In this section we present a perturbative analysis of renormalization 
constants for the local operators in eq.~(\ref{O12}) and in eq.~(\ref{O44}).  
In particular we compute the 2-loop anomalous dimension for the 
Z-factors for the two schemes discussed above. 
In bare perturbation theory the renormalization constants have an expansion 
of the form 
\begin{equation}
  \zco(g_0,a/L,x_4/L,\theta)=
 1+\sum_{k=1}^\infty \zco^{(k)}(a/L,x_4/L,\theta)\, g_0^{2k},
\end{equation}
where in the limit $a/L\to 0$
the coefficients $\zco^{(k)}$ are polynomials in
$\ln(L/a)$ of degree $k$ up to corrections of O($a/L$).
In particular the coefficient of the logarithmic divergence in
$\zco^{(1)}$ is given by the one-loop anomalous dimension $\gamma_0$,
and thus $\zco^{(1)}$ is parameterized as
\begin{equation}
\zco^{(1)} = B_\cO(\theta,x_4/L)-\gamma_0\ln(L/a)+{\rm O}(a/L)\; . 
\label{zO1}
\end{equation}
The values for the 1-loop anomalous dimension $\gamma_0$ and the 
constant piece $B_\cO(\theta,x_4/L)$ can be computed in lattice perturbation 
theory for the SF scheme following the techniques developed in 
\cite{ref:perturbative,ref:singlet}. 
For the matching of the non-perturbative lattice results and perturbation 
theory at very high energies, it is important to know also  
the 2-loop anomalous dimension $\gamma_1$\footnote{When we omit an
explicit scheme index, we always mean the SF renormalization scheme.}.
Since $\gamma_1$ is not universal, it becomes necessary to compute the 
dependence of $\gamma_1$ on the values of 
$\theta$ and $x_4/L$. In this way it becomes possible to 
control for which choice of these parameters $\gamma_1/\gamma_0$ is small
in order to have 
a good behavior of the perturbative series.                

The coefficient $\zo^{(2)}\propto \gamma_1\ln(L/a)$ 
can actually be obtained without an explicit 2-loop 
calculation in the SF scheme if the 2-loop anomalous 
dimension is already known from a different renormalization scheme \cite{ref:mass_pert}.
The formula relating the 2-loop anomalous dimensions in the SF 
and the $\overline{\mathrm{MS}}$ schemes is given by
\be
  \gamma_1 = \gamma_1^{\overline{\mathrm{MS}}} +
  2b_0 \Delta Z_{\cO}^{(1)}-\gamma_0 {\cal X}_{\rm g}^{(1)}\;.
  \label{eq:g1pert}
\ee
In eq.~(\ref{eq:g1pert}) $b_0$ is the universal 1-loop coefficient
of the $\beta$-function, $\Delta Z_{\cO}^{(1)}$ is the 1-loop
difference of the renormalization constants from a finite 
renormalization that relates two mass-independent schemes and 
${\cal X}_{\rm g}^{(1)}$ is the perturbative factor that relates 
the renormalized couplings in the two schemes considered (see the 
appendix for explicit expressions of these quantities). 

A subtlety is that the factor 
$\Delta Z_{\cO}^{(1)}$
needs to be computed in two different regularizations, in order to avoid
a calculation with the SF scheme using a dimensional regularization (DR). 

\begin{figure}
\vspace{0.0cm}
\begin{center}
\psfig{file=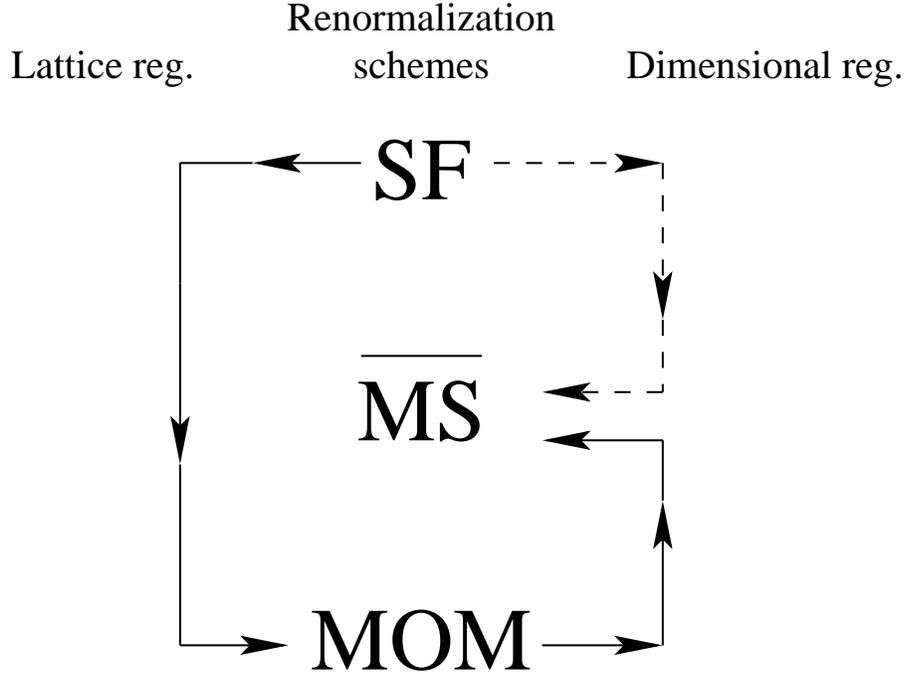, width=12cm,angle=0}
\end{center}
\caption{ \label{fig:scheme}
Graphical representation of the finite renormalization between schemes
when a different regularization is used.
}
\end{figure}

Let us explain this fact.
A matrix element $\cO_R^s$ renormalized in a certain scheme $s$ is obtained by
\be
\cO_R^s = Z_{\cO}^{s,reg} \cO_b^{reg}
\ee
where $\cO_b^{reg}$ is the bare matrix element computed within a certain
regularization $reg$ and $Z_{\cO}^{s,reg}$ is the renormalization
constant that depends on the renormalization scheme $s$ used and 
on the regularization $reg$.
Operators renormalized in two different schemes but using the same 
regularization can be related 
by a finite renormalization
\be
\cO^{s'} = \cO^s \Delta Z_{\cO}^{s'\leftarrow s}\; .
\label{eq:fin_match}
\ee
One important observation here is that $\Delta Z_{\cO}^{s'\leftarrow s}$ 
is independent from the regularization used to compute 
the renormalized matrix element and the corresponding 
renormalization constant.
This allows to relate two schemes:
in principle it is possible to compute the complete 1-loop renormalization
constant (anomalous dimension and finite part) in the SF scheme, 
using the dimensional regularization 
and then connect directly with the $\overline{\mathrm{MS}}$ scheme
(dashed line in fig.~\ref{fig:scheme}),
\be
\Delta Z_{\cO}^{SF \leftarrow \overline{\mathrm{MS}}} = 
Z_{\cO}^{SF,DR}/Z_{\cO}^{\overline{\mathrm{MS}},DR}\; .
\ee
A lattice perturbative computation will give, on the other side, 
important information
about the amount of discretization errors at 1-loop and would help
the numerical calculation. In order to avoid an additional computation in the
SF scheme one may now use 
the MOM scheme, where a complete 1-loop computation
has been done for the renormalization constant in both regularization schemes
\cite{ref:capitani}.

The desired factor $\Delta Z_{\cO}^{SF \leftarrow \overline{\mathrm{MS}}}$ 
relating the SF to the $\overline{\mathrm{MS}}$ scheme 
(full line in fig.~\ref{fig:scheme}) can now be obtained from
\be
\Delta Z_{\cO}^{SF \leftarrow \overline{\mathrm{MS}}} = 
\Delta Z_{\cO}^{SF \leftarrow MOM}
\Delta Z_{\cO}^{MOM \leftarrow \overline{\mathrm{MS}}}\; .
\ee
The two factors $\Delta Z$ are computed using different regularizations, 
namely
$\Delta Z_{\cO}^{SF \leftarrow MOM}$ is computed on the lattice
and $\Delta Z_{\cO}^{MOM \leftarrow \overline{\mathrm{MS}}}$ 
in dimensional regularization.
Using the complete one loop result \cite{ref:capitani} 
in the MOM scheme which exists in both, lattice and dimensional,
regularizations, it is then possible to compute
$\Delta Z_{\cO}^{(1)}$ and from this finally $\gamma_1$.

In table~\ref{table:gammas} we give values for $\gamma_1$ for selected values
of $\theta$ and $x_4/L=1/2$. From this table we observe that the ratio 
$\gamma_1/\gamma_0$ becomes smaller for increasing values of $\theta$ if
we choose the scheme A ($\theta=\theta'$). For scheme B ($\theta'=0$), 
we find a kind
of minimum of this ratio for $\theta\approx 1.6$. 
We repeated the analysis for the anomalous dimensions for $x_4/L=1/4$ and 
found a very similar pattern.
Thus we have a first indication for the choice of $\theta$ to be chosen. 

%
\begin{table}[t]
\centering
\begin{tabular} {|c|c|c|c|c|}
\hline
  &&&&\\[-0.5ex]
  $\theta$& $\gamma^A_1(\theta)$ & $\gamma^B_1(\theta)$ 
  & $\gamma^A_1(\theta)/\gamma_0$ & $\gamma^B_1(\theta)/\gamma_0$ \\[1ex]
\hline
  &&&&\\[-1.0ex]
$0.1$ & $\phantom{-}0.06584(1)$ & $\phantom{-}0.06551(1)$ & $\phantom{-}1.4621(6)$ 
  & $\phantom{-}1.4548(3)$
                             \\[0.5ex]
$0.4$ & $\phantom{-}0.06083(1)$ & $\phantom{-}0.05669(1)$ & $\phantom{-}1.3509(6)$ 
  & $\phantom{-}1.2590(2)$
                             \\[0.5ex]
$0.7$ & $\phantom{-}0.05196(1)$ & $\phantom{-}0.04135(1)$ & $\phantom{-}1.1539(6)$ 
  & $\phantom{-}0.9182(2)$
                             \\[0.5ex]
$1.0$ & $\phantom{-}0.04211(1)$ & $\phantom{-}0.02482(1)$ & $\phantom{-}0.9352(6)$ 
  & $\phantom{-}0.5512(2)$
                             \\[0.5ex]
$1.3$ & $\phantom{-}0.03313(1)$ & $\phantom{-}0.01029(1)$ & $\phantom{-}0.7357(6)$ 
  & $\phantom{-}0.2286(2)$
                             \\[0.5ex]
$1.6$ & $\phantom{-}0.02561(1)$ & $-0.00139(1)$ & $\phantom{-}0.5687(6)$ 
  & $-0.0308(3)$
                             \\[0.5ex]
$1.9$ & $\phantom{-}0.01951(1)$ & $-0.01051(1)$ & $\phantom{-}0.4332(6)$ 
  & $-0.2334(3)$
                             \\[0.5ex]
$2.2$ & $\phantom{-}0.01449(1)$ & $-0.01775(1)$ & $\phantom{-}0.3217(6)$ 
  & $-0.3943(3)$
                             \\[0.5ex]
$2.5$ & $\phantom{-}0.01019(1)$ & $-0.02375(1)$ & $\phantom{-}0.2263(6)$ 
  & $-0.5273(3)$
                             \\[0.5ex]
$2.8$ & $\phantom{-}0.00627(3)$ & $-0.02903(3)$ & $\phantom{-}0.1392(6)$ 
  & $-0.6446(6)$
                             \\[0.5ex]
$3.1$ & $\phantom{-}0.00231(3)$ & $-0.03412(3)$ & $\phantom{-}0.0512(6)$ 
  & $-0.7578(6)$
                             \\[0.5ex]
\hline
\end{tabular}
\caption{\footnotesize 
2-loop anomalous dimension for the $O_{44}$ representation computed in 2 different renormalization schemes
(cf. text) for several values of $\theta$ and $x_4/L=1/2$.
For $x_4/L=1/4$ the pattern is very similar.}
\label{table:gammas}
\end{table}

%

\subsection{One-loop cut-off effects}

\begin{figure}
\vspace{-2.0cm}
\begin{center}
\psfig{file=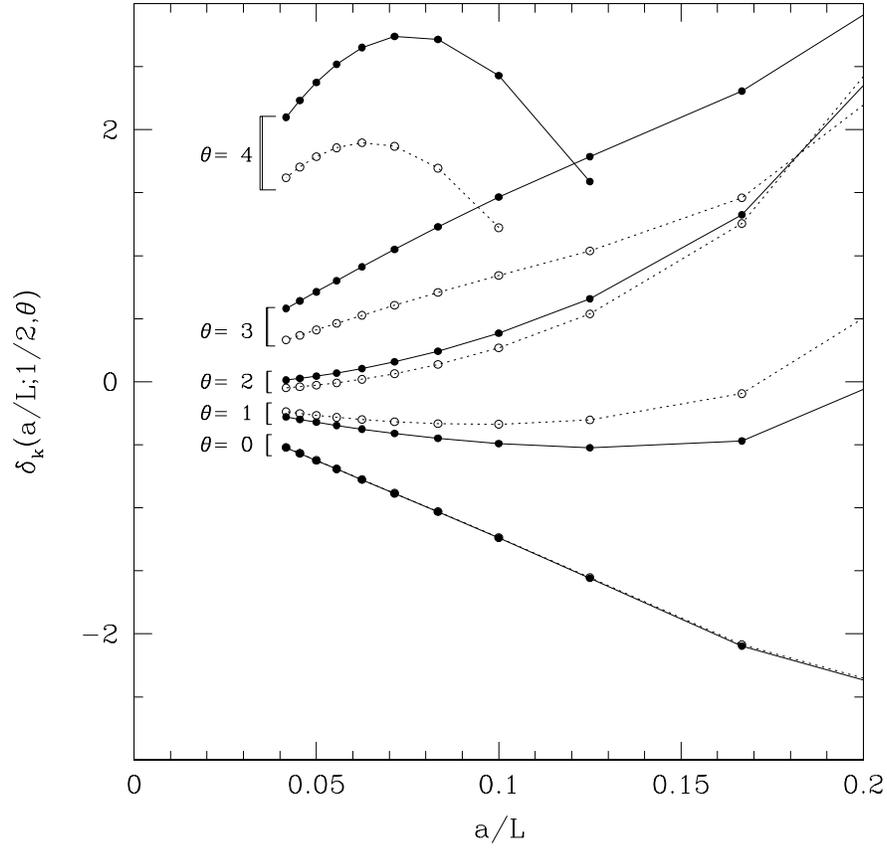, width=12cm,angle=0}
\end{center}
\caption{ \label{fig:lat_art1}
Discretization errors in the unimproved step scaling function at 1-loop order
of perturbation theory. Points are for $L/a = 6,8,\ldots, 24$ in steps of 
two.
Full symbols denote scheme A, while the open symbols denote scheme B.
}
\end{figure}

As discussed above, an essential element to obtain the non-perturbative
scale dependence of a matrix element is the step scaling function of 
eq.~(\ref{firstssf}) which describes a change from  a scale 
$L^{-1}$ to $(sL)^{-1}$, with a scale factor $s$.
Employing a lattice regularization, we define 
the lattice step scaling function  \cite{campi_bordo}
\be
\Sigma_{Z_{\cO}}(s,\bar{g}^2,a/L;x_4/L,\theta)=
\left.\frac{Z_{\cO}(g_0,sL/a;x_4/L,\theta)}
{Z_{\cO}(g_0,L/a;x_4/L,\theta)}\right\vert_{\bar{g}(L)=\mathrm{fixed}}
\label{latticessf}
\ee
where $\bar{g}(L)$ is the renormalized coupling at scale $L^{-1}$. 
The desired continuum step scaling is obtained from 
a limit procedure 
\be
\lim_{a \rightarrow 0}\Sigma_{Z_{\cal O}}(s,\bar{g}^2,a/L;x_4/L,\theta) = 
\sigma_{Z_{\cal O}}(s,\bar{g}^2;x_4/L,\theta)\; .
\label{contlimssf}
\ee
In the following, the above step scaling functions and the 
renormalized coupling are computed in the SF scheme
and the scale factor is set to $s=2$.
To one-loop order of perturbation theory we find  
\be
\Sigma_{Z_{\cO}}(2,\bar{g}^2,a/L;x_4/L,\theta) = 
1 + k(a/L;x_4/L,\theta)\bar{g}^2 + O(\bar{g}^4)
\ee
with
\be
k(a/L;x_4/L,\theta) = 
Z_{\cO}^{(1)}(a/2L;x_4/L,\theta) - Z_{\cO}^{(1)}(a/L;x_4/L,\theta)\; .
\ee
In order to see how fast the continuum limit in eq.~(\ref{contlimssf})
is approached, we define the normalized deviation from this value:        
\be
\delta_{\rm k}(a/L;x_4/L,\theta)=\frac{k(a/L;x_4/L,\theta) - k(0)}{k(0)}\; .
\label{eq:deltak}
\ee
Here $k(0;x_4/L,\theta) = -\gamma_0\ln(2)$ is the continuum limit value 
which is independent from $x_4/L$ and $\theta$.
The quantity $\delta_{\rm k}$ in eq.~(\ref{eq:deltak}) 
contains all the lattice artifacts at $O(\bar{g}^2)$.
The results for $\cO_{44}$ with $x_4/L=1/2$ are displayed in 
fig.~\ref{fig:lat_art1}. Note that we have worked 
in the unimproved theory where we expect the lattice artifacts
to decrease asymptotically with a rate proportional to $a/L$.
We show our results for $\delta_{\rm k}(a/L;x_4/L,\theta)$ for various
values of $\theta$. 
The full symbols denote results obtained in the renormalization scheme A, 
while the open symbols correspond to results
in the scheme B.

We see from fig.~\ref{fig:lat_art1} that there is no real preference 
for choosing scheme A or scheme B as far as the lattice artefacts
are concerned. On the other hand, for large values of $\theta$, the lattice
artefacts become very strong which would lead to a difficult
continuum extrapolation in the numerical simulations. It seems from
this perturbative analysis that values around $\theta=1$ should be 
a good choice for a fast convergence of perturbation theory and for 
keeping lattice artefacts well under control. 

\subsection{Signal to noise ratio}

The analysis
of the tree-level correlation function of the operator 
$\cO_{44}$ demonstrates that for $\theta=2\pi$, corresponding to the smallest
value of standard quantized momenta on the lattice,   
the value of the correlation function becomes tiny.
If this also applies in the interacting case, it may become very difficult
to obtain a reliable measurement from a numerical simulation.
On the other hand, fig.~\ref{fig:tree} shows that employing
the boundary conditions of eq.~(\ref{theta_bc}) and choosing $\theta<2\pi$ 
the situation improves considerably. 

\begin{figure}
\vspace{-2.0cm}
\begin{center}
\psfig{file=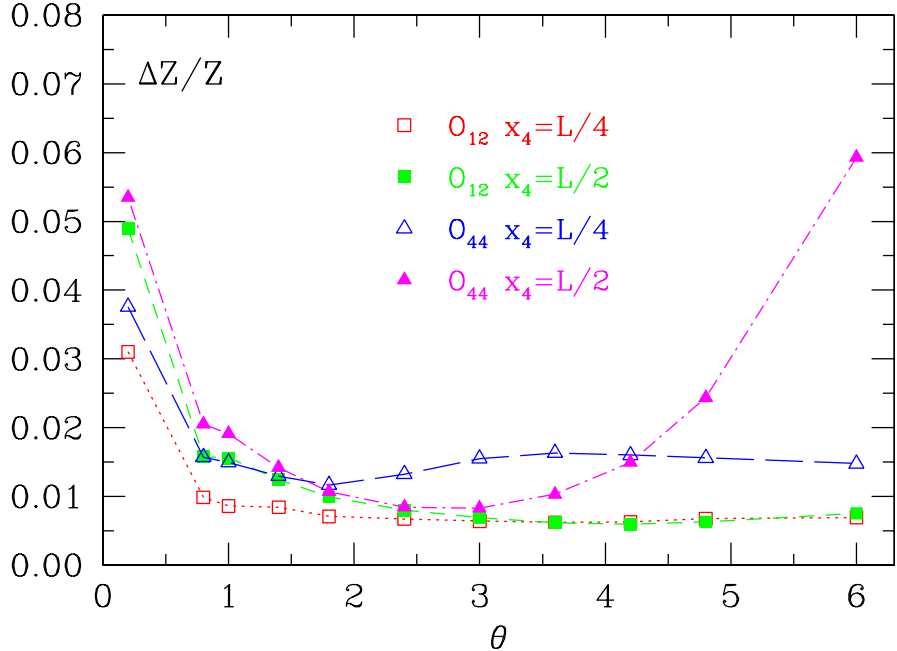, width=15cm}
\end{center}
\vspace{-6.0cm}
\caption{ \label{fig:stonoise}
$\theta$-dependence of $\Delta Z/Z$. The chosen value of 
$\kappa$ corresponds to $m_q=0$. 
The lines only connect the points to guide the eye. 
}
\end{figure}

Of course, the tree-level analysis can not give any information
on the $\theta$-dependence of the relative statistical error 
$\Delta Z/Z$ 
of a renormalization constant, which is the most important 
benchmark for a practical numerical simulation.
In order to compute 
$\Delta Z/Z$ for 
$Z_{{\cal O}_{44}}$ and $Z_{{\cal O}_{12}}$, we performed therefore a simulation
for non-perturbatively O(a)-improved (clover) fermions \cite{ref:sw} at 
$\beta=7.0203$, $\kappa=0.134707$ (corresponding to $\kappa_c$) 
on a $16^4$ lattice.
We show in fig.~\ref{fig:stonoise} the result of this 
investigation. 


Let us start the discussion on the relative error for 
$Z_{{\cal O}_{12}}$. For $\theta > 1$, the relative error appears
to be almost constant up to $\theta =2\pi$ and independent 
of choosing $x_4/L=1/2$ or $x_4/L=1/4$. This means that for
the operator ${\cal O}_{12}$ choosing a physical momentum 
(as done in our earlier work) or introducing $\theta$ 
would have led to the same quality of the final results. 

For the operator ${\cal O}_{44}$ the situation is different, however.
At $\theta=2\pi$ the relative error of $Z_{{\cal O}_{44}}$ evaluated 
at $x_4/L=1/2$ is a factor 5-6 larger than the corresponding
relative error of $Z_{{\cal O}_{12}}$. For $x_4/L=1/4$ the situation is
better, but the error is 
again about a factor 2 worse than for $Z_{{\cal O}_{12}}$
at least for $\theta=2\pi$. 
We observe a rather shallow 
minimum in $\theta$ starting around $\theta=1$ and ranging up to 
$\theta=2$ where the 
relative errors of the renormalization 
constants for both operators ${\cal O}_{12}$ and ${\cal O}_{44}$ are about the same  
and comparable to $\Delta Z/Z$ of $Z_{{\cal O}_{12}}(\theta=2\pi)$.
The case of $Z_{{\cal O}_{44}}$ is hence an example where using $\theta$ 
instead of standard lattice momenta can help substantially. 
From Fig.~\ref{fig:stonoise} the value of $\theta$ 
to be chosen for a practical
simulation can not be clearly identified.  
However, 
taken our analysis of the convergence behavior of perturbation
theory, the cutoff effects as computed in 1-loop perturbation theory
and the investigation of the signal to noise ratio together, 
we conclude 
that $\theta= 1$ should be a 
reasonable choice.                                                  

\section{Non-perturbative step scaling functions}

\begin{figure}
\begin{center}
\psfig{file=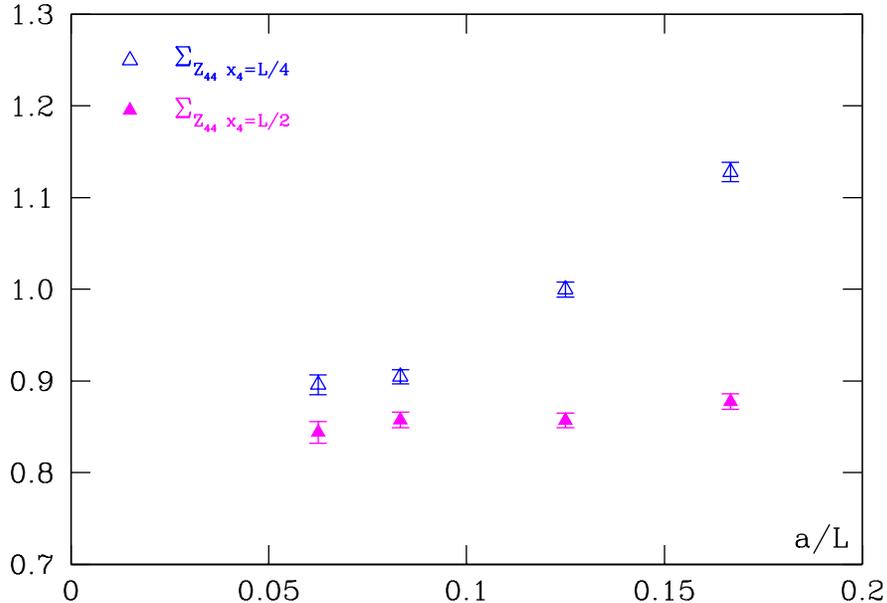, width=15cm}
\end{center}
\vspace{-6.0cm}
\caption{ \label{fig:f44T2T4}
Lattice step scaling functions $\Sigma_{Z_{44}}(a/L)$
at $x_4=L/4$ and $x_4=L/2$ at $\bar{g}^2=1.8811$.
The value of 
$\theta$ is taken from our choices in eq.~(\ref{choices}). 
}
\end{figure}

Choosing the lattice regularization discussed above, is, of course, motivated
by the advantage to perform numerical simulations. The aim of such 
simulations is to compute the lattice step scaling functions
of eq.~(\ref{latticessf}) 
non-perturbatively at non-vanishing values of the lattice spacing
and to perform the --well-defined-- continuum limit of
eq.~(\ref{contlimssf}). 
Besides the lattice step scaling function $\Sigma_{Z_{\cO}}$, we will
also consider $\Sigma_{\zbarco}$, constructed in a completely 
analogous way from the definition of the renormalization factors
in eq.~(\ref{ZO}). 
In addition, we will compute the step scaling function 
of $f_1$, eq.~(\ref{f1}),  
\begin{equation}
\label{eq:sigmas}
\sigma_{f_1}(\theta) =
                    \frac{1}{\hat{c}}\frac{\sqrt{f_1(sL,\theta)}}
                                           {\sqrt{f_1(L,\theta)}}
\end{equation}
where $\hat{c}=\sqrt{f_1^{(0)}(sL,\theta)}/\sqrt{f_1^{(0)}(L,\theta)}$ 
is the corresponding ratio at tree-level. 
The lattice version $\Sigma_{f_1}(a/L;\theta)$ of this step scaling function
is again defined in the obvious way. 
To complete the definition of the step scaling functions, certain 
choices of the values for $\theta$, $s$, $x_4/L$ and the 
quark mass $m_q$ have to be made. We fix these in the following to
\begin{equation} 
x_4/L=1/2,\; s=2,\; \theta=1,\; m_q=0\; .
\label{choices}
\end{equation} 
In order to justify our choice of $x_4/L=1/2$ we show a comparison of
the step scaling function $\Sigma_{Z_{44}}$ both at $x_4/L=1/4$
and $x_4/L=1/2$ in fig.~\ref{fig:f44T2T4}. Strong lattice artifacts are
visible for $x_4/L=1/4$. Approaching the continuum limit this effect
is considerably reduced, but it is clear that $\Sigma_{Z_{44}}$ is much
better behaved as a function of $a/L$ when $x_4/L=1/2$ is chosen, which 
allows for a reliable (linear) continuum extrapolation.
With the parameter choices in eq.~(\ref{choices}), we will in the 
following only
give the arguments of $Z_{\cal O}$ and $\sigma_{Z_{\cal O}}$ when 
they differ
from these values. In particular, with the 
choice of $s=2$ the step scaling function becomes a function of only
one scale, i.e. 
$\sigma_{Z_{\cal O}}(\mu_1/\mu_2)\rightarrow \sigma_{Z_{\cal O}}(\mu)$ and 
correspondingly the lattice step scaling function will only depend on
the scale $\mu$ (given by a fixed value of $\bar{g}(L)$) and $a/L$. 

As already discussed above, 
the continuum limit of $\Sigma_{Z_{\cal O}}(\bar{g}^2(L),a/L)$
is to be taken 
at a fixed 
value of the renormalization scale $\mu=1/L$ 
in order to obtain
the desired continuum step scaling function 
$\sigma_{Z_{\cal O}}(\bar{g}^2(L))$.
Fixing the scale is realized by fixing 
the renormalized coupling $\bar{g}^2(L)$,
which in turn can be achieved by
changing the value of the bare coupling as the value of the 
lattice spacing (and correspondingly $a/L$) is 
decreased. Using the SF scheme, the matching points of the 
bare coupling $g_0$ to keep 
$\bar{g}^2(L)$ fixed when the lattice spacing is sent to zero 
can be found in \cite{alfa_mass}. 
Of course, the values of $\Sigma_{Z_{\cal O}}(\bar{g}^2(L),a/L)$ 
themselves have to 
be computed in a numerical simulation on lattices with increasing
number of points $N^4=(L/a)^4$ as $a\rightarrow 0$.

As in our previous work, we will choose two different 
lattice discretizations, standard Wilson fermions and 
non-perturbatively O(a)-improved (clover) fermions to compute
the step scaling function $\Sigma_{Z_{\cal O}}(\bar{g}^2(L),a/L)$.            
Since we want to employ a massless renormalization scheme and have 
to stay therefore at zero quark, we 
take from \cite{alfa_mass} the values 
of $\kappa_c$ for clover fermions to stay in the massless limit. For the
case of Wilson fermions, we determined $\kappa_c$ ourselves (see table \ref{sim_par}).
%
%
\begin{table}
\begin{center}
\begin{tabular}{|c|c|r|l|l|}
\hline
     $\overline{g}_{SF}^2$ 
  &  $L/a$ 
  &  $\beta \quad\quad$ 
  &  $\;\;\kappa_c [Wilson]$ 
  &  $\;\;\kappa_c [Clover]$
  \\
\hline
      $~ 0.8873$ 
  &   $6$ 
  &   $~ 10.7503 ~$ 
  &   $~ 0.134696(7)$ 
  &   $~ 0.130591(4)$ 
  \\
      $$ 
  &   $8$ 
  &   $~ 11.0000 ~$ 
  &   $~ 0.134548(6)$ 
  &   $~ 0.130439(3)$ 
  \\
      $$ 
  &   $12$ 
  &   $~ 11.3384 ~$ 
  &   $~ 0.134277(5)$ 
  &   $~ 0.130251(2)$ 
  \\
      $$ 
  &   $16$ 
  &   $~ 11.5736 ~$ 
  &   $~ 0.134068(6)$ 
  &   $~ 0.130125(2)$ 
  \\
\hline
      $1.0989$ 
  &   $6$ 
  &   $~ 9.5030 ~$ 
  &   $~ 0.136520(5)$ 
  &   $~ 0.131514(5)$ 
  \\
      $$ 
  &   $8$ 
  &   $~ 9.7500 ~$ 
  &   $~ 0.136310(3)$ 
  &   $~ 0.131312(3)$ 
  \\
      $$ 
  &   $12$ 
  &   $~ 10.0577 ~$ 
  &   $~ 0.135949(4)$ 
  &   $~ 0.131079(3)$ 
  \\
      $$ 
  &   $16$ 
  &   $~ 10.3419 ~$ 
  &   $~ 0.135572(4)$ 
  &   $~ 0.130876(2)$ 
  \\
\hline
      $1.3293$ 
  &   $6$ 
  &   $~ 8.6129 ~$ 
  &   $~ 0.138346(6)$ 
  &   $~ 0.132380(6)$ 
  \\
      $$ 
  &   $8$ 
  &   $~ 8.8500 ~$ 
  &   $~ 0.138057(4)$ 
  &   $~ 0.132140(5)$ 
  \\
      $$ 
  &   $12$ 
  &   $~ 9.1859 ~$ 
  &   $~ 0.137503(2)$ 
  &   $~ 0.131814(3)$ 
  \\
      $$ 
  &   $16$ 
  &   $~ 9.4381 ~$ 
  &   $~ 0.137061(4)$ 
  &   $~ 0.131589(2)$ 
  \\
\hline
      $1.5533$ 
  &   $6$ 
  &   $~ 7.9993 ~$ 
  &   $~ 0.140003(11)$ 
  &   $~ 0.133118(7)$ 
  \\
      $$ 
  &   $8$ 
  &   $~ 8.2500 ~$ 
  &   $~ 0.139588(8)$ 
  &   $~ 0.132821(5)$ 
  \\
      $$ 
  &   $12$ 
  &   $~ 8.5985 ~$ 
  &   $~ 0.138847(6)$ 
  &   $~ 0.132427(3)$ 
  \\
      $$ 
  &   $16$ 
  &   $~ 8.8323 ~$ 
  &   $~ 0.138339(7)$ 
  &   $~ 0.132169(3)$ 
  \\
\hline
      $1.8811$ 
  &   $6$ 
  &   $~ 7.4082 ~$ 
  &   $~ 0.142145(11)$ 
  &   $~ 0.133961(8)$ 
  \\
      $$ 
  &   $8$ 
  &   $~ 7.6547 ~$ 
  &   $~ 0.141572(9)$ 
  &   $~ 0.133632(6)$ 
  \\
      $$ 
  &   $12$ 
  &   $~ 7.9993 ~$ 
  &   $~ 0.140597(6)$ 
  &   $~ 0.133159(4)$ 
  \\
      $$ 
  &   $16$ 
  &   $~ 8.2415 ~$ 
  &   $~ 0.139900(6)$ 
  &   $~ 0.132847(3)$ 
  \\
\hline
      $2.1000$ 
  &   $6$ 
  &   $~ 7.1214 ~$ 
  &   $~ 0.143416(11)$ 
  &   $~ 0.134423(9)$ 
  \\
      $$ 
  &   $8$ 
  &   $~ 7.3632 ~$ 
  &   $~ 0.142749(9)$ 
  &   $~ 0.134088(6)$ 
  \\
      $$ 
  &   $12$ 
  &   $~ 7.6985 ~$ 
  &   $~ 0.141657(6)$ 
  &   $~ 0.133599(4)$ 
  \\
      $$ 
  &   $16$ 
  &   $~ 7.9560 ~$ 
  &   $~ 0.140817(7)$ 
  &   $~ 0.133229(3)$ 
  \\
\hline
      $2.4484$ 
  &   $6$ 
  &   $~ 6.7807 ~$ 
  &   $~ 0.145286(11)$ 
  &   $~ 0.134994(11)$ 
  \\
      $$ 
  &   $8$ 
  &   $~ 7.0197 ~$ 
  &   $~ 0.144454(7)$ 
  &   $~ 0.134639(7)$ 
  \\
      $$ 
  &   $12$ 
  &   $~ 7.3551 ~$ 
  &   $~ 0.143113(6)$ 
  &   $~ 0.134141(5)$ 
  \\
      $$ 
  &   $16$ 
  &   $~ 7.6101 ~$ 
  &   $~ 0.142107(6)$ 
  &   $~ 0.133729(4)$ 
  \\
\hline
      $2.7700$ 
  &   $6$ 
  &   $~ 6.5512 ~$ 
  &   $~ 0.146825(11)$ 
  &   $~ 0.135327(12)$ 
  \\
      $$ 
  &   $8$ 
  &   $~ 6.7860 ~$ 
  &   $~ 0.145859(7)$ 
  &   $~ 0.135056(8)$ 
  \\
      $$ 
  &   $12$ 
  &   $~ 7.1190 ~$ 
  &   $~ 0.144299(8)$ 
  &   $~ 0.134513(5)$ 
  \\
      $$ 
  &   $16$ 
  &   $~ 7.3686 ~$ 
  &   $~ 0.143113(?)$ 
  &   $~ 0.134114(3)$ 
  \\
\hline
      $3.4800$ 
  &   $6$ 
  &   $~ 6.2204 ~$ 
  &   $~ 0.149685(15)$ 
  &   $~ 0.135470(15)$ 
  \\
      $$ 
  &   $8$ 
  &   $~ 6.4527 ~$ 
  &   $~ 0.148391(9)$ 
  &   $~ 0.135543(9)$ 
  \\
      $$ 
  &   $12$ 
  &   $~ 6.7750 ~$ 
  &   $~ 0.146408(7)$ 
  &   $~ 0.135121(5)$ 
  \\
      $$ 
  &   $16$ 
  &   $~ 7.0203 ~$ 
  &   $~ 0.145025(8)$ 
  &   $~ 0.134707(4)$ 
  \\
\hline
\end{tabular}
\caption{Simulation Parameters}
\label{sim_par}
\end{center}
\end{table}

Simulations have been performed on APEmille machines using even--sized
lattices ranging from $6^4$ to $32^4$. We recall that for both actions
we used the unimproved operator: as a consequence, the results at
finite values of the lattice spacing are affected by lattice artefacts
of O$(a)$, whose precise form depends upon the choice of the lattice
action.
                                                             
For the inversion of the Dirac operator we used the implementation
\cite{ssor} of the SSOR-preconditioned BiCGStab inverter
\cite{bicgstab}.
The gauge fields were generated with a hybrid of Cabibbo--Marinari
heat--bath and over--relaxation updates: we decided to keep the number
of over--relaxation steps between a single heat-bath one proportional
to $L/a$ (in practice $N_{\rm or} = \frac{1}{2}L/a$). A full update
sweep is defined as one heat--bath step followed by $N_{\rm or}$
steps: in order to statistically decorrelate the gauge configurations
used to take measurements of our observables, we allowed for a number
of full sweeps between measurements ranging from 20 (lattices up to $L/a =
16$) to 70 ($L/a = 32$).

As usual in this kind of finite--size--step simulations, the signal
over noise ratio deteriorates for larger values of $\gbar^2$, in the
fully non--perturbative regime. In order to maintain constant, up to a
certain degree, the relative statistical errors on our observables, we
are obliged to increase the statistics. For example, in the clover case and
on $L/a = 32$ lattices, the number of measurements has to grow from around
300 at the smallest values of $\gbar^2$ up to order 700 at the largest
value of the renormalized coupling. Accordingly, for lattices
with a smaller number of space--time points, the number of
measurements grows from 500 to 2000. The pattern for the Wilson case is
similar, albeit with an overall smaller statistical sample.
                                                         

\begin{figure}
\begin{center}
\psfig{file=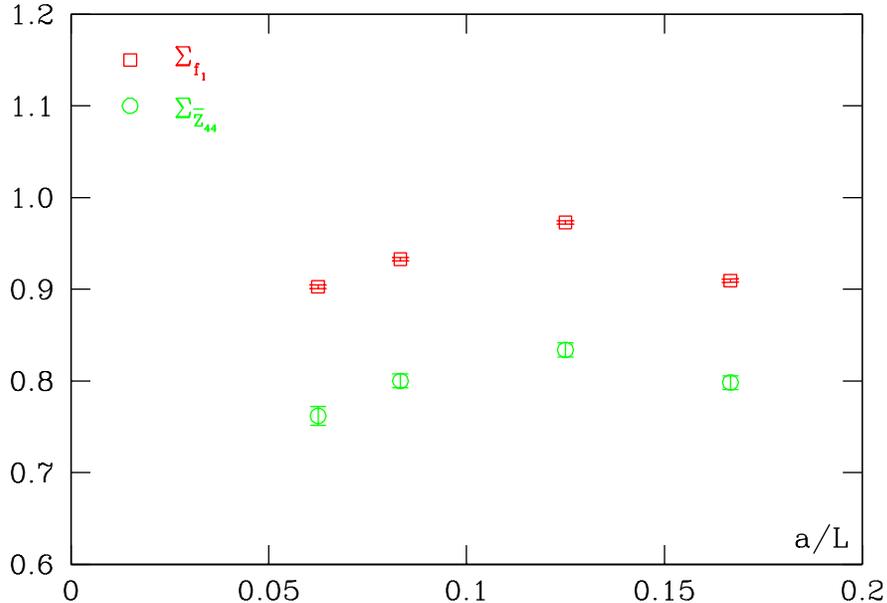, width=15cm}
\end{center}
\vspace{-6.0cm}
\caption{ \label{fig:f1f44}
Lattice step scaling functions $\Sigma_{f_1}(\bar{g}^2(L),a/L)$ and 
$\Sigma_{\bar{Z}_{44}}(\bar{g}^2(L),a/L)$ at $\bar{g}^2=1.8811$. 
The values of 
$\theta$ and $x_4/L$ are taken from our choices in eq.~(\ref{choices}). 
}
\end{figure}

In Fig.~\ref{fig:f1f44} we show separately 
the lattice step scaling functions $\Sigma_{\bar{Z}_{44}}(\bar{g}^2(L),a/L)$ 
and 
$\Sigma_{f_1}(\bar{g}^2(L),a/L)$ at four values of $a/L$ for $\bar{g}^2$ fixed.  
Both step scaling functions show rather large
lattice artefacts. For $\Sigma_{f_1}$ this behavior is in sharp 
contrast to the case of choosing $\theta=0$, where the $a/L$ dependence 
of $\Sigma_{f_1}$ is basically flat, at least in the O(a)-improved
theory \cite{ref:non-pert,ref:universal}. 
This flat behavior led us, in our earlier work,
to the conclusion to perform the continuum limit of 
$\Sigma_{\bar{Z}_{12}}$ and $\Sigma_{f_1}$ separately and to compute 
the desired continuum step scaling function
$\sigma_{Z_{12}}=\sigma_{\bar{Z}_{12}}/\sigma_{f_1}$ after the 
continuum limit of $\Sigma_{\bar{Z}_{12}}$ and 
$\Sigma_{f_1}$ had been performed. 

The example of the step scaling functions, 
shown in Fig.~\ref{fig:f1f44}, indicates
that for $\theta$ non-vanishing the strategy might be different. 
Indeed, we found in the analysis of our data --for $\theta=1$--
that in the ratios 
\begin{equation}
\Sigma_{\bar{Z}_{44}}(\bar{g}^2(L),a/L)/\Sigma_{f_1}(\bar{g}^2(L),a/L)
\label{ratio1}
\end{equation}
and 
\begin{equation}
\Sigma_{\bar{Z}_{12}}(\bar{g}^2(L),a/L)/\Sigma_{f_1}(\bar{g}^2(L),a/L) 
\label{ratio2}
\end{equation}
the lattice artefacts essentially cancel. This cancellation happens 
for clover as well as for Wilson fermions. 
We decided therefore that for our choice of $\theta=1$ to first compute 
\begin{equation}
\Sigma_{Z}(\bar{g}^2(L),a/L)=
\Sigma_{\bar{Z}}(\bar{g}^2(L),a/L)/\Sigma_{f_1}(\bar{g}^2(L),a/L) 
\label{ratio3}
\end{equation}
at a given value 
of $a/L$ and then perform the continuum limit. 


\begin{figure}
\vspace{-2.0cm}
\begin{center}
\psfig{file=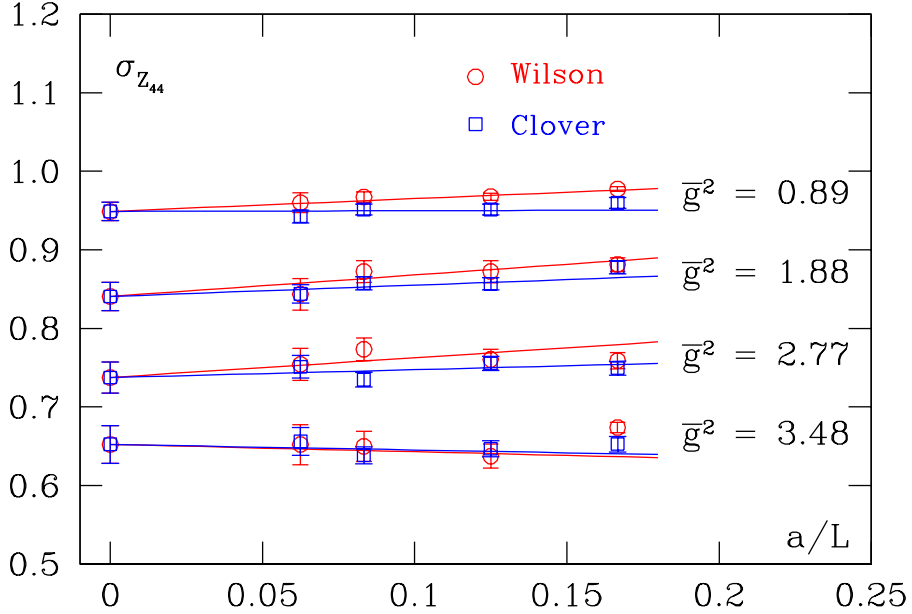,width=15cm}
\end{center}
\vspace{-6.0cm}
\caption{ \label{fig:s44_cont}
Examples of the continuum limit of step scaling functions
for both Wilson and clover fermions. The (constrained) 
fit performed is 
linear, taking the 3 data points with smallest values of
$a/L$. We also indicate the values of the running coupling 
that correspond to these step scaling functions.
}
\end{figure}

In Fig.~\ref{fig:s44_cont} we show examples of the continuum extrapolations 
of $\Sigma_{{\cal O}_{44}}$ for both discretization, Wilson and 
clover fermions. It can be observed that in both cases 
the continuum extrapolation is linear in the lattice spacing when the 
data points with the three smallest values of $a/L$ are taken. In the 
plot we show already a constraint fit, demanding that both 
set of lattice results extrapolate to the same value of the 
continuum step scaling function. 
We obtain a very similar plot for $\Sigma_{{\cal O}_{12}}$. 
The numerical results of the continuum extrapolation as well as for the
four lattice spacings are summarized in table \ref{res_sigma_Z12} for the
${\cal O}_{12}$ and in table \ref{res_sigma_Z44} for the ${\cal O}_{44}$ operator.

One subtlety, we want to mention, is that even in the continuum limit
at fixed scale, 
the step scaling functions of ${\cal O}_{44}$ and ${\cal O}_{12}$ do not assume
the same value. The reason is that both operators belong to two different
representations and in the here adopted {\em finite volume} 
renormalization scheme they are therefore also part of the 
precise definition of the scheme. 
It is hence not possible, to constrain the fits even more by 
demanding that $\Sigma_{{\cal O}_{44}}$ and $\Sigma_{{\cal O}_{12}}$
converge to the same value of the continuum step scaling function.

\section{Running and invariant step scaling function}

At this stage, we can leave the lattice and discuss {\em continuum}
quantities only. The only reminder that we have performed a lattice 
calculation is that 
we will stay in 
the somewhat unusual, finite volume SF renormalization scheme. 
The simulations described 
in the previous section provide us with 
the continuum step scaling functions $\sigma_{Z_{44}}$ and 
$\sigma_{Z_{12}}$  at a number of values for the running
coupling $0.8 < \bar{g}^2< 3.5$, corresponding to a wide 
range of scales, $500 \;\mathrm{MeV} < \mu < 100 \;\mathrm{GeV}$.  


\begin{figure}
\vspace{-2.0cm}
\begin{center}
\psfig{file=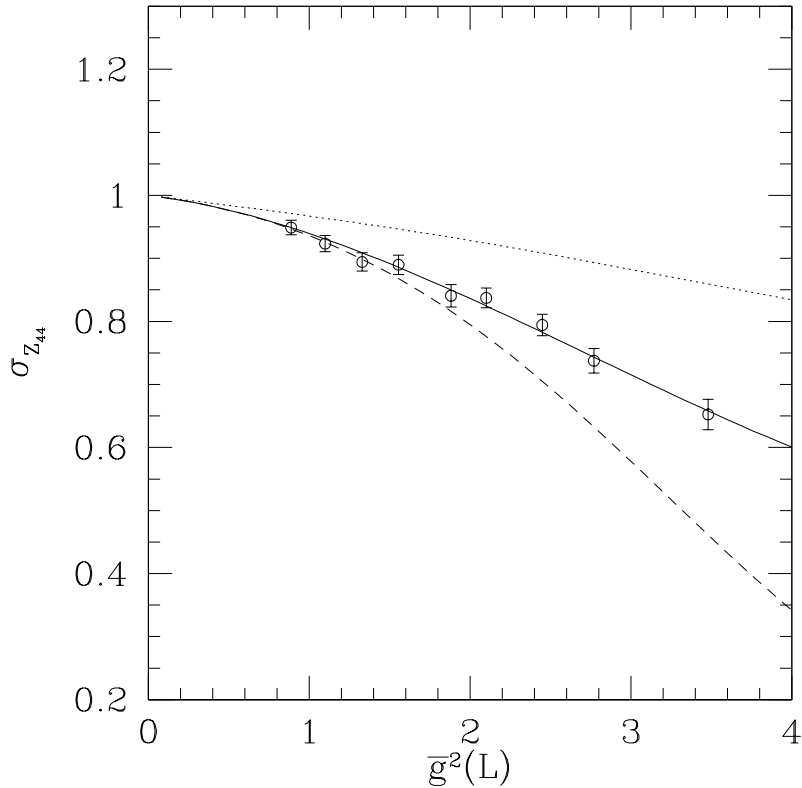,width=11cm,height=11cm}
\end{center}
\caption{ \label{fig:sigma44}
Continuum step scaling function $\sigma_{Z_{44}}$. The solid line
represents the parameterization eq.~(\ref{fitsigma}), the dotted line
represents the the 1-loop, the dashed line the 2-loop predictions
from perturbation theory. 
}
\end{figure}

In Fig.~\ref{fig:sigma44} we show  
the continuum step scaling function $\sigma_{Z_{44}}$
as a function of the running coupling (and hence the scale).
The dotted curve is the 1-loop perturbative 
analysis for the step scaling function, whereas
the dashed curve is a 2-loop result, taking the 2-loop anomalous 
dimension in the SF scheme from its relation to the 2-loop 
anomalous dimension in the $\overline{\mathrm{MS}}$-scheme discussed
in section 2.1. The solid line is a fit 
to the data according to the formula
\begin{equation}
\sigma_{Z_{44}}=1-\gamma_0 \ln(2) g_0^2 + c_1 g_0^4 + c_2 g_0^6\; .
\label{fitsigma}
\end{equation}
In eq.~(\ref{fitsigma}) we have taken the known 1-loop ($\gamma_0$) 
and fitted only the coefficients $c_1,c_2$ which we find to be
$c_1=-0.0334(50),\; c_2=0.0041(18)$.
It can be seen from the figure that 1-loop perturbation theory is not a good 
description of the data at all. Even 2-loop perturbation theory represents
the data only up to $\bar{g}^2=2$, or so. 


\begin{figure}
\vspace{-2.0cm}
\begin{center}
\psfig{file=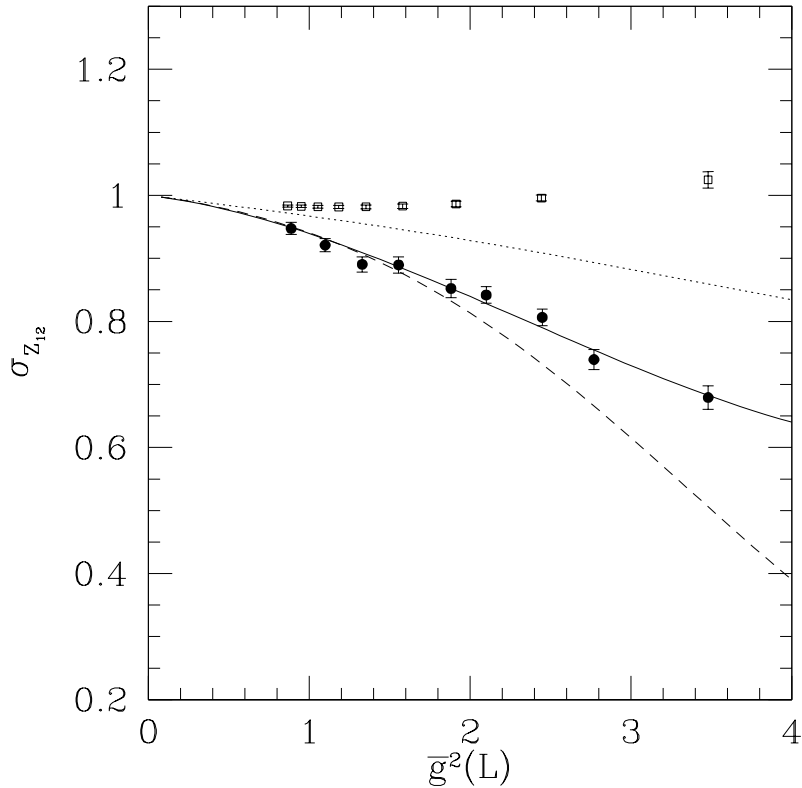,width=11cm,height=11cm}
\end{center}
\caption{ \label{fig:sigma12}
Continuum step scaling function $\sigma_{Z_{12}}$. The solid line
represents the parameterization eq.~(\ref{fitsigma}), the dotted line
represents the the 1-loop, the dashed line the 2-loop predictions
from perturbation theory.                                   
We also show the values of $\sigma_{Z_{12}}$ for 
$\theta=2\pi$ (open symbols) corresponding to a minimal 
physical momentum.
}
\end{figure}

Fig.~\ref{fig:sigma12} is the same as 
fig.~\ref{fig:sigma44} for the operator ${\cal O}_{12}$. 
In this case we find the coefficients $c_1=-0.0343(41),\; c_2=0.0049(14)$
for the parameterization of eq.~(\ref{fitsigma}).
In addition, we show the values of the step scaling function 
$\sigma_{Z_{12}}^{p}$ as obtained in our earlier work, i.e. using 
$\theta=0$ and a momentum $p=2\pi/L$ (open symbols).
Clearly, the scale dependence
of $\sigma_{Z_{12}}(\theta=1)$ is much stronger than for 
$\sigma_{Z_{12}}^{p}$.

The final result of this paper is the 
(ultra-violet) invariant step scaling function
for the operators ${\cal O}_{44}$ and ${\cal O}_{12}$
as they are needed to describe the running of the 
physical matrix element in other schemes than the SF one.
We adopt here the definition of the invariant step scaling function
as given in \cite{ref:invariant}:
\begin{equation}
  \EuFrak{S}_{\rm INV}^{\rm UV,SF}(\mu_0) = \sigma ( \mu/\mu_0,\bar{g}^2(L_0) )\cdot
                     (\bar{g}^2(L))^{-\gamma_0/2b_0} \exp\left\{
                     -\int_0^{\bar{g}(\mu)}
                     dg\left[\frac{\gamma(g)}{\beta(g)} 
                     - \frac{\gamma_0}{b_0g}\right]\right\}\; .
\label{eq:step_running_inv}
\end{equation}
The $\beta$- and $\gamma$-functions in 
eq.~(\ref{eq:step_running_inv}) will be taken to a given 
order in perturbation theory in the SF scheme. 
A scheme dependence is still manifest in $\EuFrak{S}_{\rm INV}$
through the appearance of the infra-red scale ($\mu_0=L_0^{-1}$).
We thus add a superscript to indicate that the invariant
step scaling function is computed within the Schr\"odinger functional scheme.

For large enough scales in the ultra-violet, it is expected that the running of 
the step scaling functions, as computed non-perturbatively
above, will match the perturbative running and hence that
$\EuFrak{S}_{\rm INV}$ will become independent from the scale.
In order to test at which scales $\EuFrak{S}_{\rm INV}$ becomes 
constant, 
we have to evolve the step scaling functions starting from our most 
non-perturbative scale $\mu_0 = (1/2L_{\rm{max}})$, implicitly given by
$\bar{g}^2(2L_{\rm{max}}) = 3.48$. One finds \cite{ref:Lmax_r0}
$2L_{\rm{max}} = (1.476r_0)$, where $r_0 \approx 0.5\mathrm{fm}$.
The step scaling functions $\sigma_{Z_{\cal O}}$ are only computed at 
certain values of $\bar{g}^2(L)$. Subsequent values of  $\bar{g}^2(L)$
do, however, not correspond to a scale change by a factor of two as
needed. 
In order to find the precise evolution by steps of two an interpolation 
in $\bar{g}^2(L)$ has to be performed using 
the parameterization of eq.~(\ref{fitsigma}). The corresponding values 
of the step scaling function can be found in table~\ref{interpol}. 

The error of the so obtained values of $\sigma_{Z_{\cal O}}$ is evaluated by
a standard error propagation taking the correlation of the 
fit parameters into account through the covariance matrix (cov), i.e. 
\begin{equation}
(\Delta \sigma_{Z_{\cal O}}) = \sqrt{\frac{\partial \sigma_{Z_{\cal O}}}{\partial c_1}^2(\Delta c_1)^2
                    + \frac{\partial \sigma_{Z_{\cal O}}}{\partial c_2}^2(\Delta c_2)^2
                    + 2 \frac{\partial \sigma_{Z_{\cal O}}}{\partial c_1}
                \cdot \frac{\partial \sigma_{Z_{\cal O}}}{\partial c_2}
                \cdot \mathrm{cov}_{12}} + \frac{\partial \sigma_{Z_{\cal O}}}{\partial \bar{g}^2} \Delta \bar{g}^2\; .                         
\label{deltafitsigma}
\end{equation}
As an aside we mention that we also have seen that the errors coming from 
the uncertainty in $\bar{g}^2(L)$ are not negligible, 
and they are included in the values given in table \ref{interpol}.
In evaluating $\EuFrak{S}_{\rm INV}$ in eq.~(\ref{eq:step_running_inv})
we have taken the 3-loop $\beta$-function and the 2-loop $\gamma$-function.

\begin{figure}
\vspace{-2.0cm}
\begin{center}
\psfig{file=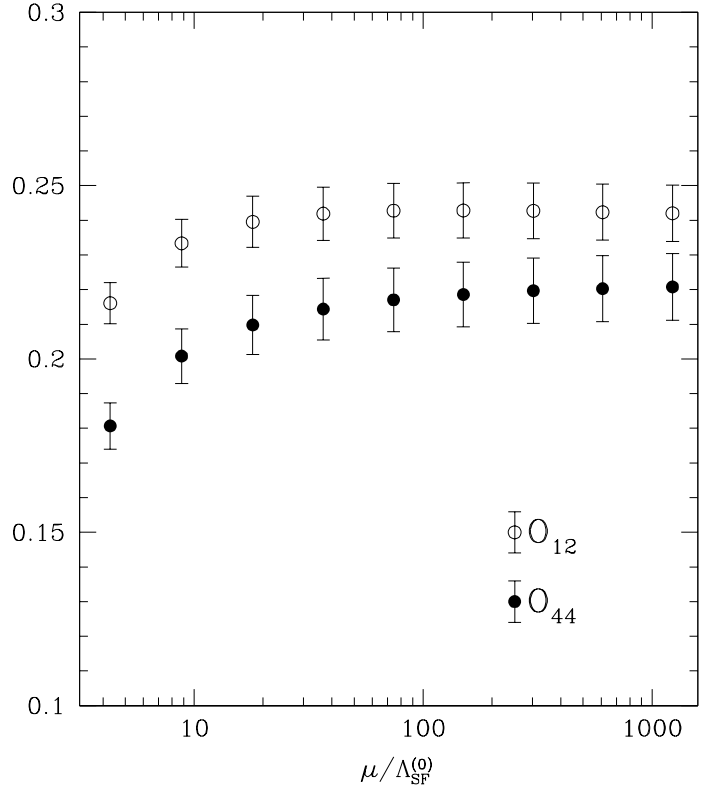,width=11cm,height=11cm}
\end{center}
\caption{ \label{fig:uvsigma}
The RG invariant step scaling function for the operator 
${\cal O}_{44}$ (full symbols) and the operator ${\cal O}_{12}$ (open symbols).
}
\end{figure}

In Fig.~\ref{fig:uvsigma} we show $\EuFrak{S}_{\rm INV}$ as a function of 
$\mu/\Lambda_\mathrm{SF}^{(0)}$ with 
$\Lambda_\mathrm{SF}^{(0)}\approx 120 \;\mathrm{MeV}$ being the 
$\Lambda$-parameter in the quenched approximation 
in the SF scheme.
For the operator ${\cal O}_{44}$ (full symbols) the invariant step scaling 
function is constant for, say, $\mu/\Lambda_\mathrm{SF}^{(0)} > 50$ 
indicating that contact with perturbation theory can safely be made.
In the same figure we also show the value of the invariant 
step scaling function for ${\cal O}_{12}$ which shows a very similar behavior. 
The fact that both invariant step scaling functions assume different
values indicates again that the two different operators define
two different renormalization schemes, at least in a finite volume
scheme like the SF. The difference will only disappear in the 
physical renormalization group invariant matrix element.

In order to obtain finally the values of the invariant step scaling functions,
we have taken the values at the corresponding largest scale: 
\begin{equation}
\EuFrak{S}_{\rm INV,{\cal O}_{12}}^{UV,SF}(\mu_0) = 0.242(8),\; 
\EuFrak{S}_{\rm INV,{\cal O}_{44}}^{UV,SF}(\mu_0) = 0.221(9)\; .
\end{equation}
As a comparison we quote the value of 
$\EuFrak{S}_{\rm INV,{\cal O}_{12}}^{UV,SF}(\mu_0) = 1.11(4)$ as obtained 
at $\theta=0$ and $p=2\pi/L$ in our previous work. 


\vspace{2.0cm}

\section{Conclusion}

In this paper we have demonstrated how the generalized boundary conditions
advocated in ref.~\cite{campi_bordo} can to utilized 
to define continuous external momenta as often needed in the lattice
simulations when operators with derivatives are considered. 
We investigated the particular example of a scale 
dependent renormalization constant for renormalizing twist-2 
non-singlet quark operators corresponding to moments of
parton distribution functions.

Although we demonstrated the usefulness for such momentum definitions 
for a renormalization constant, or more specific its step scaling function,
it is clear that in principle the same method also applies 
to matrix elements themselves, as long as no real physical 
momentum transfer 
is involved. The feasibility of this approach has to be tested, however,
in a real numerical benchmark simulation.

For the particular example at hand, we could determine the renormalization
constant for a certain lattice representation of a continuum operator 
that determines the average (quark) momentum in a hadron. 
Such a calculation would have been very difficult by using standard, quantized 
lattice momenta as usually taken. As a result, we could determine
the RG invariant step scaling function for two different lattice
representations of the same continuum operator. 
In this way, we could eliminate a small systematic uncertainty that
plagued our earlier work.

The idea of the present work was to show the practicability and usefulness 
of the generalized momentum definition on a practical example. 
The values of the RG invariant step scaling we have computed are 
essential ingredients for the RG invariant matrix elements which 
allow a comparison with experimental data or global fits. 
Results for these physical matrix elements,
in which we are finally interested in, will be reported in a 
forthcoming publication. 

\subsection*{Acknowledgments}
We thank S. Capitani, R. Sommer and S. Sint for many useful discussions.       
The computer center at NIC/DESY Zeuthen provided the necessary technical 
help and the computer resources. 
This work was supported by
the EU IHP Network on Hadron Phenomenology from Lattice QCD
and by the DFG Sonderforschungs\-bereich/Transregio SFB/TR9-03.

%
%
\begin{table}
\begin{center}
\begin{tabular}{|c|c|l|l|l|l|}
\hline
$\bar{g}_{SF}^2$ & $a/L$ & $\Sigma_{Z_{12}} [Wilson]$ & $\Sigma_{Z_{12}} [Clover]$ &
$~~~~~~\sigma_{Z_{12}}$ & $\chi^2/dof$\\
\hline
~      &  1/6 &~  0.9937(25)  &~  0.9682(60) &~             &~       \\
0.8873 &  1/8 &~  0.9787(34)  &~  0.9616(59) &~ 0.9475(96)  &~ 0.32  \\
~      & 1/12 &~  0.9640(55)  &~  0.9571(58) &~             &~       \\
~      & 1/16 &~  0.9689(94)  &~  0.9555(71) &~             &~       \\
\hline                                                              
~      &  1/6 &~  0.9899(39)  &~  0.9540(58) &~             &~       \\
1.0989 &  1/8 &~  0.9727(55)  &~  0.9591(62) &~ 0.9211(105) &~ 0.23  \\
~      & 1/12 &~  0.9608(67)  &~  0.9446(61) &~             &~       \\
~      & 1/16 &~  0.9480(91)  &~  0.9368(74) &~             &~       \\
\hline                                                              
~      &  1/6 &~  0.9771(42)  &~  0.9360(69) &~             &~       \\
1.3293 &  1/8 &~  0.9680(61)  &~  0.9365(67) &~ 0.8905(120) &~ 0.79  \\
~      & 1/12 &~  0.9339(81)  &~  0.9305(66) &~             &~       \\
~      & 1/16 &~  0.9230(117) &~  0.9157(81) &~             &~       \\
\hline                                                              
~      &  1/6 &~  0.9716(48)  &~  0.9153(64) &~             &~       \\
1.5553 &  1/8 &~  0.9545(77)  &~  0.9102(67) &~ 0.8896(129) &~ 0.26  \\
~      & 1/12 &~  0.9287(92)  &~  0.9068(63) &~             &~       \\
~      & 1/16 &~  0.9133(132) &~  0.9031(86) &~             &~       \\
\hline                                                              
~      &  1/6 &~  0.9607(68)  &~  0.8984(75) &~             &~       \\
1.8811 &  1/8 &~  0.9259(104) &~  0.8778(70) &~ 0.8522(144) &~ 1.07  \\
~      & 1/12 &~  0.9060(103) &~  0.8781(72) &~             &~       \\
~      & 1/16 &~  0.8682(153) &~  0.8640(93) &~             &~       \\
\hline                                               
~      &  1/6 &~  0.9627(68)  &~  0.8700(71) &~             &~       \\
2.1000 &  1/8 &~  0.9107(74)  &~  0.8592(70) &~ 0.8419(131) &~ 1.01  \\
~      & 1/12 &~  0.8981(101) &~  0.8558(72) &~             &~       \\
~      & 1/16 &~  0.8827(102) &~  0.8368(99) &~             &~       \\
\hline                                                        
~      &  1/6 &~  0.9467(67)  &~  0.8217(75) &~             &~       \\
2.4484 &  1/8 &~  0.8811(85)  &~  0.8223(77) &~ 0.8064(133) &~ 0.78  \\
~      & 1/12 &~  0.8589(136) &~  0.8053(77) &~             &~       \\
~      & 1/16 &~  0.8519(85)  &~  0.8116(107)&~             &~       \\
\hline                                                        
~      &  1/6 &~  0.9309(75)  &~  0.7864(75) &~             &~       \\
2.7700 &  1/8 &~  0.8759(94)  &~  0.7838(78) &~ 0.7395(157) &~ 0.77  \\
~      & 1/12 &~  0.8385(102) &~  0.7593(75) &~             &~       \\
~      & 1/16 &~  0.8225(152) &~  0.7583(118)&~             &~       \\
\hline                                                        
~      &  1/6 &~  0.9063(50)  &~  0.7176(83) &~             &~       \\
3.4800 &  1/8 &~  0.8203(110) &~  0.6899(86) &~ 0.6793(187) &~ 0.56  \\
~      & 1/12 &~  0.7846(144) &~  0.6761(89) &~             &~       \\
~      & 1/16 &~  0.7604(191) &~  0.6851(136)&~             &~       \\
\hline
\end{tabular}
\caption{Results for the lattice step scaling function $\Sigma_{Z_{12}}$
and the combined continuum extrapolation $\sigma_{Z_{12}}$ of Wilson and
Clover data with the three smallest lattice spacings.}
\label{res_sigma_Z12}
\end{center}
\end{table}

%
%
\begin{table}
\begin{center}
\begin{tabular}{|c|c|l|l|l|l|}
\hline
$\bar{g}_{SF}^2$ & $a/L$ & $\Sigma_{Z_{44}} [Wilson]$ & $\Sigma_{Z_{44}} [Clover]$ &
$~~~~~~\sigma_{Z_{44}}$ & $\chi^2/dof$\\
\hline
~      &  1/6 &~  0.9776(30)  &~  0.9596(68) &~             &~       \\
0.8873 &  1/8 &~  0.9677(44)  &~  0.9513(68) &~ 0.9489(115) &~ 0.49  \\
~      & 1/12 &~  0.9671(65)  &~  0.9514(68) &~             &~       \\
~      & 1/16 &~  0.9601(125) &~  0.9425(81) &~             &~       \\
\hline                                                       
~      &  1/6 &~  0.9577(49)  &~  0.9375(66) &~             &~       \\
1.0989 &  1/8 &~  0.9560(69)  &~  0.9421(71) &~ 0.9234(128) &~ 0.50  \\
~      & 1/12 &~  0.9505(85)  &~  0.9420(71) &~             &~       \\
~      & 1/16 &~  0.9330(123) &~  0.9290(86) &~             &~       \\
\hline                                                        
~      &  1/6 &~  0.9402(54)  &~  0.9184(77) &~             &~       \\
1.3293 &  1/8 &~  0.9416(81)  &~  0.9275(77) &~ 0.8943(145) &~ 1.40  \\
~      & 1/12 &~  0.9250(104) &~  0.9322(78) &~             &~       \\
~      & 1/16 &~  0.9118(153) &~  0.9037(93) &~             &~       \\
\hline                                                        
~      &  1/6 &~  0.9062(63)  &~  0.8908(71) &~             &~       \\
1.5553 &  1/8 &~  0.9163(100) &~  0.8977(75) &~ 0.8900(154) &~ 1.60  \\
~      & 1/12 &~  0.9140(117) &~  0.9084(74) &~             &~       \\
~      & 1/16 &~  0.8834(163) &~  0.8873(103)&~             &~       \\
\hline                                                        
~      &  1/6 &~  0.8813(88)  &~  0.8777(84) &~             &~       \\
1.8811 &  1/8 &~  0.8723(140) &~  0.8569(80) &~ 0.8407(178) &~ 0.54  \\
~      & 1/12 &~  0.8723(139) &~  0.8577(84) &~             &~       \\
~      & 1/16 &~  0.8434(199) &~  0.8439(117)&~             &~       \\
\hline                                                        
~      &  1/6 &~  0.8698(88)  &~  0.8354(80) &~             &~       \\
2.1000 &  1/8 &~  0.8467(97)  &~  0.8399(82) &~ 0.8373(157) &~ 1.75  \\
~      & 1/12 &~  0.8545(133) &~  0.8425(84) &~             &~       \\
~      & 1/16 &~  0.8590(129) &~  0.8181(112)&~             &~       \\
\hline                                                        
~      &  1/6 &~  0.8197(88)  &~  0.8016(89) &~             &~       \\
2.4484 &  1/8 &~  0.7935(119) &~  0.7885(91) &~ 0.7944(169) &~ 0.55  \\
~      & 1/12 &~  0.7942(186) &~  0.7921(94) &~             &~       \\
~      & 1/16 &~  0.7823(115) &~  0.8036(127)&~             &~       \\
\hline                                                        
~      &  1/6 &~  0.7589(103) &~  0.7494(88) &~             &~       \\
2.7700 &  1/8 &~  0.7606(129) &~  0.7556(90) &~ 0.7376(197) &~ 1.22  \\
~      & 1/12 &~  0.7736(143) &~  0.7347(92) &~             &~       \\
~      & 1/16 &~  0.7542(203) &~  0.7512(143)&~             &~       \\
\hline                                                        
~      &  1/6 &~  0.6739(66)  &~  0.6528(99) &~             &~       \\
3.4800 &  1/8 &~  0.6376(153) &~  0.6469(104)&~ 0.6525(241) &~ 0.34  \\
~      & 1/12 &~  0.6500(193) &~  0.6385(110)&~             &~       \\
~      & 1/16 &~  0.6522(256) &~  0.6563(178)&~             &~       \\
\hline
\end{tabular}
\caption{Results for the lattice step scaling function $\Sigma_{Z_{44}}$
and the combined continuum extrapolation $\sigma_{Z_{44}}$ of Wilson and
Clover data with the three smallest lattice spacings.}
\label{res_sigma_Z44}
\end{center}
\end{table}

%
%
\begin{table}
\begin{center}
\begin{tabular}{|l|l|l|}
\hline
~~~ $\bar{g}_{SF}^2$ &~~~~~ $\sigma_{Z_{12}}$ &~~~~~ $\sigma_{Z_{44}}$\\
\hline
~ 3.480     &~  0.6826(169) &~  0.6579(216)  \\
~ 2.454(18) &~  0.7893(83)  &~  0.7822(100)  \\
~ 1.918(18) &~  0.8486(76)  &~  0.8458(90)   \\
~ 1.584(18) &~  0.8840(67)  &~  0.8828(79)   \\
~ 1.353(18) &~  0.9071(58)  &~  0.9066(68)   \\
~ 1.184(17) &~  0.9231(50)  &~  0.9229(58)   \\
~ 1.053(15) &~  0.9348(42)  &~  0.9348(49)   \\
~ 0.950(14) &~  0.9436(37)  &~  0.9437(42)   \\
~ 0.865(13) &~  0.9505(32)  &~  0.9506(37)   \\
\hline
\end{tabular}
\caption{Interpolated values of the step scaling function $\sigma_{Z_{12}}$
and $\sigma_{Z_{44}}$ for the scale change by a factor two.}
\label{interpol}
\end{center}
\end{table}

\newpage

\begin{appendix}

\section*{Appendix A}

In this appendix we recall some basic properties of the renormalization group
functions.
For small couplings the $\beta$-function and the anomalous dimension ($\gamma$-function),
defined by
\be\meqalign{
\beta(g) &=& \mu \frac{\partial}{\partial \mu} g(\mu) \cr
\gamma(g) &=& \mu \frac{\partial}{\partial \mu} \log Z_{\cO}(\mu,g)
}
\label{RG_eq}
\ee
have asymptotic expansions of the form
\be\meqalign{
  \beta(g) &\buildrel{g}\rightarrow0\over\sim
            & -g^3\sum_{k=0}^\infty b_k g^{2k},\cr 
  \gamma(g)  &\buildrel{g}\rightarrow0\over\sim
            & -g^2\sum_{k=0}^\infty \gamma_k g^{2k}.
} 
\ee
One finds that $b_0$, $b_1$ and $\gamma_0$ 
are the same in all the schemes (these are the ``universal" coefficients), 
while all other coefficients are scheme dependent. 
The universal coefficients are given by (with $C_F=(N_c^2-1)/2N_c$ and $N_c$ the number of colors)
\begin{eqnarray}
  b_0 &=& \bigl\{\frac{11}{3}N_c-\frac{2}{3}\Nf\bigr\}(4\pi)^{-2},\\[1ex]
  \gamma_0 &=& \frac{16}{3}C_F(4\pi)^{-2} ,\\[1ex]
  b_1 &=& \bigl\{\frac{34}{3}N_c^2-(\frac{13}{3}N_c-N_c^{-1})\Nf\bigr\}(4\pi)^{-4}.
  \label{gamma1msbar}
\end{eqnarray}
Any two mass independent renormalization schemes can be related
by a scale change and a finite parameter renormalization of the form
\begin{eqnarray}
  \mu'&=&c\mu,\qquad c > 0,\\[1ex]
  \bar{g}'&=&\bar{g} \sqrt{{\cal X}_{\rm g}(\bar{g})},\\[1ex]
  Z'&=&Z [\Delta Z_{\cO}(\bar{g})], 
\label{eq:fin_ren}
\end{eqnarray}
where $c$ is just a change of scale between the 2 schemes and one could obviously choose 
also $c=1$.
${\cal X}_{\rm g}$ and $\Delta Z_{\cO}$ are expanded according to
\be
  {\cal X}_{\rm g}(\bar{g})\buildrel{\bar{g}}\rightarrow0\over\sim
  1+\sum_{k=1}^{\infty}{\cal X}_{\rm g}^{(k)}\bar{g}^{2k}.
\ee
\be
  \Delta Z_{\cO}(\bar{g})\buildrel{\bar{g}}\rightarrow0\over\sim
  1+\sum_{k=1}^{\infty}\Delta Z_{\cO}^{(k)}\bar{g}^{2k}.
\ee
The invariance of a physical observable under such a change of parameters, 
gives a relation between the renormalization group functions, $\beta$ and $\gamma$, 
in the 2 schemes. In particular we have 
\be
  \gamma_1 = \gamma_1^{\overline{\mathrm{MS}}} +
  2b_0 \Delta Z_{\cO}^{(1)}-\gamma_0 {\cal X}_{\rm g}^{(1)}.
  \label{eq:gamma1pert}
\ee
From ref. \cite{ref:coupl_pert} we have
\be
  {\cal X}^{(1)}_{\rm g}=-{1\over 4\pi}(c_{1,0}+c_{1,1}\Nf),
  \label{Xg}
\ee
with
\be
  c_{1,0}=1.25563(4),\qquad  c_{1,1}=0.039863(2).
\ee
From the perturbative results in \cite{ref:capitani} (section 2.1) 
it is possible to obtain $\Delta Z_{\cO}^{(1)}$.

\end{appendix}

\def\NPB #1 #2 #3 {Nucl.~Phys.~{\bf#1} (#2)\ #3}
\def\NPBproc #1 #2 #3 {Nucl.~Phys.~B (Proc. Suppl.) {\bf#1} (#2)\ #3}
\def\PRD #1 #2 #3 {Phys.~Rev.~{\bf#1} (#2)\ #3}
\def\PLB #1 #2 #3 {Phys.~Lett.~{\bf#1} (#2)\ #3}
\def\PRL #1 #2 #3 {Phys.~Rev.~Lett.~{\bf#1} (#2)\ #3}
\def\PR  #1 #2 #3 {Phys.~Rep.~{\bf#1} (#2)\ #3}

\def\etal{{\it et al.}}
\def\ibid{{\it ibid}.}

\newpage

\def\NPB #1 #2 #3 {Nucl.~Phys.~{\bf#1} (#2)\ #3}
\def\NPBproc #1 #2 #3 {Nucl.~Phys.~B (Proc. Suppl.) {\bf#1} (#2)\ #3}
\def\PRD #1 #2 #3 {Phys.~Rev.~{\bf#1} (#2)\ #3}
\def\PLB #1 #2 #3 {Phys.~Lett.~{\bf#1} (#2)\ #3}
\def\PRL #1 #2 #3 {Phys.~Rev.~Lett.~{\bf#1} (#2)\ #3}
\def\PR  #1 #2 #3 {Phys.~Rep.~{\bf#1} (#2)\ #3}

\def\etal{{\it et al.}}
\def\ibid{{\it ibid}.}

\end{document}